\documentstyle[fleqn,11pt]{article}


\begin{document}

\setlength{\baselineskip}{0.67cm}

\title{Bounded Concurrent Timestamp Systems Using Vector Clocks\thanks{
This research was supported in parts by the Netherlands Organization for
Scientific Research (NWO) under Contract Number NF 62-376
(NFI project ALADDIN), EU fifth framework project QAIP, IST--1999--11234,
the NoE QUIPROCONE IST--1999--29064, the ESF QiT Programmme,
and the EU Fourth Framework BRA NeuroCOLT II Working Group EP 27150.
This research started when the first author was visiting
the Department of Computer Science, Utrecht University, the Netherlands
with support from the Netherlands Organization for Scientific Research (NWO)
under Contract Number NF 62-376 (NFI project ALADDIN:
{\em Algorithmic Aspects of Parallel and Distributed Systems}), 
and continued while he was at the Tata Institute of Fundamental Research, 
Mumbai, India.
%
% Parts of this research were done when the first author was visiting
% the Department of Computer Science, Utrecht University, the Netherlands
% with support from the Netherlands Organization for Scientific Research (NWO)
% under Contract Number NF 62-376 (NFI project ALADDIN:
% {\em Algorithmic Aspects of Parallel and Distributed Systems}),
% and the Tata Institute of Fundamental Research, Mumbai, India.
% The second author was supported in part by 
% the Netherlands Organization for Scientific Research (NWO)
% under Contract Number NF 62-376 (NFI project ALADDIN, 
% EU fifth framework project QAIP, IST--1999--11234, the NoE QUIPROCONE IST--1999--29064,
% the ESF QiT Programmme, and the EU Fourth Framework BRA
 % NeuroCOLT II Working Group
% EP 27150.
}
}
\author{S.~Haldar\thanks{Present address:
TimesTen Performance Software, 1991 Landings Drive,
Mountain View, CA~094043, USA.}\\[1mm]
Bell Laboratories \\
                        600 Mountain Avenue, \\
                        Murray Hill, NJ 07974, USA. \\
                        haldar@cs.mun.ca \\
\and P.M.B. Vit\'{a}nyi \\[1mm]
Centrum voor Wiskunde en Informatica \\
Kruislaan 413, 1098 SJ Amsterdam       \\
The Netherlands \\
Email: paulv@cwi.nl
}
\date{}

\maketitle
\newcommand{\ra}{\longrightarrow}
\newcommand{\LA}{\Longrightarrow}
\newcommand{\RA}[1]{\Longrightarrow_{#1}}
\newcommand{\cf}{\stackrel{cf}{\ra}}
\newcommand{\da}{\parbox[c]{22pt}{\begin{picture}(22,0)(0,0) \multiput(4,0)(5,0){2}{\line(1,0){3}}
                                                  \put(14,0){\vector(1,0){4}}
                       \end{picture}}}

\newtheorem{theorem}{Theorem}
\newtheorem{lemma}{Lemma}
\newtheorem{assertion}{Assertion}
\newtheorem{claim}{Claim}
\newtheorem{corollary}{Corollary}
\newtheorem{proposition}{Proposition}

\newcommand{\buff}{\mbox{{\em buff\/}}}
\newcommand{\RC}{\mbox{{\em r\/}}}
\newcommand{\WC}{\mbox{{\em w\/}}}
\newcommand{\seen}{\mbox{{\em seen\/}}}
\newcommand{\copyc}{\mbox{{\em copyc\/}}}
\newcommand{\copybuff}{\mbox{{\em copybuff\/}}}
\newcommand{\prevc}{\mbox{{\em prevc\/}}}

\newcommand{\wpk}{\mbox{$W_{p}^{[k]}$}}
\newcommand{\wpki}{\mbox{$W_{p}^{[k+1]}$}}
\newcommand{\wpkj}{\mbox{$W_{p}^{[k+2]}$}}
\newcommand{\wqt}{\mbox{$W_{q}^{[t]}$}}
\newcommand{\wqti}{\mbox{$W_{q}^{[t+1]}$}}
\newcommand{\wqtj}{\mbox{$W_{q}^{[t+2]}$}}
\newcommand{\ril}{\mbox{$R_{i,p}^{[l]}$}}
\newcommand{\rilj}{\mbox{$R_{i,j}^{[l]}$}}
\newcommand{\rili}{\mbox{$R_{i}^{[l+1]}$}}
\newcommand{\rpi}{\mbox{$r_{i,p}$}}
\newcommand{\rqj}{\mbox{$r_{q,j}$}}

\newcommand{\wpi}{\mbox{$w_{p,i}$}}
\newcommand{\wqj}{\mbox{$w_{q,j}$}}

\newcommand{\wpj}{\mbox{$w_{p,j}$}}

\newcommand{\rjm}{\mbox{$R_{j}^{[m]}$}}

\renewcommand{\index}{\mbox{$index$}}
\begin{abstract}
Shared registers are basic objects used as communication mediums
in asynchronous concurrent computation. 
A concurrent timestamp system is a higher typed communication object,
and has been shown to be a powerful tool to solve many 
concurrency control problems.
It has turned out to be possible to
construct such higher typed objects from 
primitive lower typed ones. The next step is 
to find efficient constructions. 
We propose a very efficient wait-free construction of bounded  
concurrent timestamp systems from
1-writer multireader registers. This finalizes, corrects, and extends,
a preliminary bounded multiwriter construction 
proposed by the second author in 1986. That work partially initiated the
current interest in wait-free concurrent objects, and 
introduced a notion of discrete vector
clocks in distributed algorithms.
\end{abstract}

\vspace{2mm}
\noindent{\bf Categories and Subject Descriptors:}
B.3.2 [{\bf Memory Structures}]: Design Styles --- {\em shared memory};
B.4.3 [{\bf Input/Output and Data Communications}]: 
      Interconnections (subsystems) --- {\em asynchronous/synchronous operation};
D.1.3 [{\bf Programming Techniques}]: Concurrent Programming;
D.4.1 [{\bf Operating Systems}]:
Process Management --- {\em concurrency, 
multiprocessing/multiprogramming\/};
D.4.4 [{\bf Operating Systems}]: Communications Management --- 
{\em buffering\/};

\vspace{2mm}
\noindent{\bf General Terms:} Algorithms, Theory, Verification

\vspace{2mm}
\noindent{\bf Additional Key Words and Phrases:}
 Concurrent reading while writing; label; nonatomic operation execution; 
operation --- read and write, labeling and scan; operation execution; 
shared variable --- safe, regular and atomic; 
timestamp system, traceability, vector clock, wait-freedom.

\section{Introduction} \label{sec1}
Consider a system of asynchronous processes that communicate among themselves
by executing read and write operations on a set of 
shared variables (also known
as shared {\em registers}) only.
The system has no global clock or any synchronization primitives.
Every shared variable is associated with a process (called {\em owner\/}) which writes it and
the other processes may read it.
An execution of a write (read) operation on a shared variable will
be referred to as a {\em Write\/}  ({\em  Read\/}) on that variable.
A Write on a shared variable puts a value from a pre determined finite domain into the
variable, and a Read   reports a value from the domain.
A process that writes (reads) a variable is called a {\em writer\/} ({\em reader\/}) of the
variable.

\noindent {\bf Wait-free shared variable:}
We want to construct shared variables in which 
the following two properties hold.
(1)~Operation executions are not necessarily atomic, 
that is, they are not indivisible,
and (2)~every operation finishes its execution 
within a bounded number of its
own steps, irrespective of the presence of other operation executions and
their relative speeds. That is, operation executions are {\em wait-free}.
These   two properties give rise to a classification of shared variables,
depending on their output characteristics.
Lamport \cite{lamp86} distinguishes three 
categories for 1-writer shared variables,
using a precedence relation on operation executions defined as follows:
for operation executions $A$ and $B$, $A$ {\em precedes\/} $B$, denoted $A\ra
B$, if $A$ finishes before $B$ starts; $A$ and $B$ {\em overlap\/} if neither
$A$ precedes $B$ nor $B$ precedes $A$.
In 1-writer variables, all the Writes are totally ordered by ``$\ra$''.
The three categories of 1-writer shared variables defined by Lamport are the following.
\begin{enumerate}
\item A {\em safe\/} variable is one in which a Read not overlapping any Write
returns the most recently written value.
A Read that overlaps a Write may return any value from the domain of the variable.
 
\item A {\em regular\/} variable is a safe variable in which
a Read that overlaps one or more Writes returns either the value
of the most recent Write preceding the Read or of one of the overlapping
Writes.
 
\item An {\em atomic\/} variable is a regular variable in which
the Reads and Writes behave as if they occur in some total order
which is an extension of the precedence relation.
\end{enumerate}
 
A shared variable is {\em boolean\/}\footnote{Boolean variables are referred to as {\em bits}.} 
or {\em multivalued\/} depending upon whether it can hold only two or more than two values.
 
\noindent {\bf Multiwriter shared variable:}
%Paul: I rewrote the text below.
A multiwriter shared variable is one that can be written and read
(concurrently) by many processes.
Lamport  \cite{lamp86} constructed a shared variable
that could be written by one process and read by one other
process, but he did not consider constructions
of shared variables with more than one writer or reader.
Vit\'{a}nyi and Awerbuch \cite{vit86} were the first to construct
an atomic multiwriter shared variable from 1-writer variables.
They propose
two constructions: one from 1-writer multireader shared
variables using bounded control information
that turned out to be incorrect \cite{vit87} 
(just regular and not atomic as claimed), 
and the other from
1-writer 1-reader variables using unbounded control information. 
The latter construction is correct. 
It is made bounded in \cite{li89},
yielding one of the most optimal implementations
that are currently known. (In this paper we correct and extend
the first construction to obtain an efficient version of the more general
notion of bounded concurrent timestamp system as defined below.)
 Related work is
\cite{abra95,bloom87,burn87,hald91,hald95,hald96,isra92a,kiro87,lamp86,LV92,li89,pete83,pete87,scha88,sing87,wolf87,vidya90}.
In particular,
it is now possible to construct bounded multiwriter atomic variables from
1-writer 1-reader safe bits.
See \cite{li89}, and the last section of this paper, 
for a brief history of the subject.
 
\noindent {\bf Timestamp system:}
In a multiwriter shared variable it is only required that every process keeps track
of which process wrote last. There arises the general question whether every process
can keep track of the order of the last Writes by all processes.  
This idea was formalized by Israeli and Li \cite{isra93}. They introduced
and analyzed the notion of
{\em timestamp system\/} as an abstraction of such
a higher typed communication medium. In a timestamp system
every process owns an {\em object\/}, 
an abstraction of a set of shared variables.
One of the requirements of the system is to determine the temporal order in
which the objects are written.
For this purpose, each object is given a {\em label\/} (also refer to as {\em
timestamp}) which indicates the latest (relative) time when it has been 
written by its owner process.
The processes assign labels to their respective objects in such a way that the 
labels reflect the real-time order in which they are written to. 
These systems must support two operations, namely {\em labeling\/} and {\em
scan}.
A labeling operation execution (Labeling, in short) assigns a new label to an
object, and a scan operation execution (Scan, in short) enables a process to
determine the ordering in which all the objects are written,
that is, it returns a set of labeled-objects ordered temporally.
We are concerned with those systems where operations can be
executed {\em concurrently}, in an overlapped fashion.
Moreover, operation executions must be {\em wait-free}, that is, each
operation execution will take a bounded
number of its own steps (the number of
accesses to the shared space), irrespective of the presence of other operation
executions and their relative speeds.

Wait-free constructions of concurrent timestamp systems (CTSs, in short)
have been 
shown to be a powerful tool for solving concurrency control problems such as
{\em fcfs}-mutual exclusion \cite{dijk65,lamp74}, multiwriter multireader shared variables
\cite{vit86}, probabilistic consensus 
\cite{abra88,chor87}, {\em fcfs} $l$-exclusion
\cite{fisc79} by synthesizing a ``wait-free
clock'' to sequence the actions in a concurrent system.

Here, we are interested in constructing concurrent timestamp systems using 
1-writer shared variables.
It is not difficult to construct a timestamp system if the shared space is
unbounded (there is no limit on the size of some shared variables).
The problem gets much harder for bounded (shared space) systems.
A {\em bounded timestamp system\/} is a timestamp system with a 
finite set of bounded size labels.
In the rest of the paper, unless stated otherwise, by a timestamp system 
we mean a wait-free bounded concurrent timestamp system.

Israeli and Li \cite{isra93} 
constructed a bit-optimal bounded timestamp system 
for sequential operation executions.
The {\em concurrent} case of bounded 
timestamp system is harder and the first generally
accepted solution is due to Dolev and Shavit
\cite{dole89}.
Their construction is of the type as in \cite{isra93}
and uses shared variables of size $O(n)$, where $n$ is the
number of processes in the system.
Each Labeling requires $O(n)$ steps, and each Scan $O(n^2\log n)$ steps.
In their construction, no Scan writes any shared variables: 
It is a `pure' reading operation execution.
(But, by the theorem of Lamport \cite[page~91]{lamp86}, all such
constructions become de facto impure if we break them down
to the lowest level of
system building.)
Following Dolev and Shavit, several researchers have come up with 
other constructions.
Israeli and Pinhasov  
\cite{isra92} use shared  variables of
size $O(n^2)$;  Labeling and Scan require $O(n)$ steps.
Gawlik, Lynch and Shavit \cite{gawl92} use shared  variables of
size $O(n^2)$;  Labeling and Scan access $O(n\log n)$ shared variables.
In \cite{dwork92}, Dwork and Waarts introduce
a powerful communication abstraction 
called ``traceable use abstraction''  to recycle
values of shared variables.
They demonstrate the usefulness of the abstraction by constructing a CTS,
borrowing the basic ideas and techniques
from \cite{vit86} for recycling private values.
Their construction requires shared  variables of size
$O(n\log n)$; Labeling and Scan require $O(n)$ steps.
Later, they along with Herlihy and Plotkin \cite{dwork92a} 
propose a construction using shared
variables of size $O(n)$; Labeling and Scan access $O(n)$ shared variables.
Unlike the Israeli-Li and Dolev-Shavit constructions, 
Scans in other proposed constructions are not pure;
they write a lot of shared space.

\noindent  {\bf Our result and related work:}
Among the constructions mentioned above, 
the one of  Dwork and Waarts \cite{dwork92}
is relatively simple and efficient 
as well\footnote{We find it is the easiest one to
understand; also see comments in \cite{yako93} by Yakovlev.}. 
They introduce ``traceable use abstraction'' to bound the size of labels.
Like in \cite{vit86}, each label is a vector of $n$ private values, 
one for each of $n$ processes.
Using a strategy similar to, and extending, \cite{vit86}, 
the abstraction helps each process to keep track of its private values
that are in use in the system.
At any point in time, a process can use only a bounded number of private
values of another process.
Exploiting that feature, the abstraction helps 
in bounding the set of private values needed.
The labels are read by executing a {\em traceable-read\/} function, and
written by executing a {\em traceable-write\/} procedure.
When the traceable-read function is executed to read a 
label,  the executing process explicitly 
informs all other processes which of their
private values it is going to use.
A process can find which of its private values are in use by other
processes even if the values propagate through these processes in tandem
one after another.
To determine which of its private values are currently not in use, a process
executes a {\em garbage collection\/} routine. This routine helps processes to
safely recycle their respective private values that are not in use. 
These three routines are at the heart of implementing the traceable use abstraction.
Dwork and Waarts \cite{dwork92} have shown how these routines are used in 
constructing a bounded concurrent timestamp system. 
The most intricate among these routines is the garbage collection,
whose time complexity is $O(n^2)$ that could be, though nonstandard, uniformly amortized over $O(n^2)$
labeling operation executions.
To achieve this, each process needs to maintain a private,
separate, pool of $22n^2$  private values.
The costliest part of their construction is the use 
of multireader `order' variables.
The construction uses, for each process, $\Theta (n)$ sets of $22n$-many
 1-writer $n$-reader atomic variables of size
$\Theta(n\log n)$ bits each. 
Let us roughly estimate their space complexity at the fundamental level, i.e., at the level
of 1-writer 1-reader safe bits.
(To implement a 1-writer $n$-reader atomic
variable of size $m$ bits, the constructions  in \cite{lamp86,vidya90}
together require $3mn$ 1-writer 1-reader safe bits, $2n$ 1-writer 1-reader
atomic bits and one 1-writer $n$-reader atomic bit.
Each 1-writer 1-reader atomic bit can be implemented from $O(1)$
1-writer 1-reader safe bits \cite{hald91,lamp86,tromp89,vidya96}.
A 1-writer $n$-reader atomic bit can be implemented from
$O(n^2)$ safe bits \cite{hald91}.
Thus, we require a total of $3mn + O(n^2)$ 1-writer 1-reader safe bits
to implement a 1-writer $n$-reader atomic variables of size $m$ bits.)
% It is now well known that implementing a 1-writer $n$-reader atomic 
% variables of size $m$ bits requires at least $3mn$ 1-writer 1-reader safe bits, plus $O(n^2)$ control 
% bits \cite{hald91,pete83,vidya90}.) 
Thus, there is a need of at least  $\Omega(n^4\log n)$ bits 
at the fundamental level just for the order variables in each process.
Consequently, we need at least $\Omega(n^5\log n)$ 1-writer 1-reader
safe bits for all order variables of all processes.
In addition, there are other shared variables for the processes.

The bounded multiwriter shared variable construction 
of Vit\'{a}nyi and Awerbuch
\cite{vit86}, while falling short of the claimed atomicity \cite{vit87},
%have introduced many techniques that were used later for wait-free computing,
has brought into prominence many techniques 
that were used later in wait-free computing.
An example is
the idea of a label as a vector of $n$ individual clocks.\footnote{
The concept of vector clock is used in many areas of distributed
computing, all in related contexts, to keep track of execution evolution
in distributed systems.
(Cf.~The articles by Mattern\cite{matt89,matt92}.)}
(In \cite{vit86}, vector entries are called `tickets'.)
Even better, it turns out that 
the corrected version presented here 
suffices to implement the higher communication object type of bounded CTS. 
The current paper is the final version of the
pioneering preliminary \cite{vit86}, and its correction \cite{hald93}.
Dwork and Waarts \cite{dwork92}, without stating this explicitly, 
used the idea of (bounded) vector clocks and 
other techniques introduced in \cite{vit86},
and hence their solution
bears a close resemblance to the construction proposed here (and, in fact,
to other constructions \cite{pete87,scha88} based on \cite{vit86}).
On the other hand, our
construction uses some ideas from their traceable use abstraction. 
We observe that in CTSs the propagation of private values is restricted to
only one level of indirection, and not to arbitrary levels.
Consequently, the propagation of private values
can be tracked down
by their respective owner processes with relative ease.
And, the one level indirect propagation of private values by other processes
need not be informed to the original owner of these private values.
Thus, one doesn't need the 
complete power of the traceable use abstraction for
constructing a CTS.
In our construction, we use less powerful traceable-read and traceable-write.
But, we prefer to use the same function/procedure names of \cite{dwork92}
just keep conformity with the literature.
We do not require a garbage collection routine, 
thereby simplifying the proposed CTS construction and its
correctness proof considerably. 
When a process executes the traceable-read function, it does not explicitly
inform the other processes which of their private values it is going to use.
On the other hand, the executers of the traceable-write procedure
correctly find which private values of which processes are in use in the system.
Another important point is that, in our construction, a 
Scan writes a limited amount of information,
only $O(n)$ 1-writer 1-reader bits.
Also, each local pool of private values contains fewer than $2n^2$ values.
We use a total of $n^2$ $O(n\log n)$ bit size 1-reader 1-writer regular order variables, requiring a total of
$O(n^3\log n)$ safe 1-reader 1-writer bits at the fundamental level.
Both the scan and labeling operation executions require $O(n)$ steps in
terms of the shared variables used.
But in our construction, a Scan reads at most 
($n-1$) 1-writer 1-reader regular order variables,
whereas in their construction it is ($2n-2$) 1-writer $n$-reader atomic ones.
Thus, at the fundamental level they scan order of magnitude more
bits than we do. 

Our construction is not optimal in terms of the usage of shared space
(Cf.~Table~\ref{tab1} in Section~\ref{sec5}). It is perhaps possible
to use a bounded set of global values and to recycle
them instead of using private values. 
Recycling of global values could lead to an optimal
construction.

The remainder of this paper is organized as follows.
Section~\ref{sec2} discusses the system model 
and presents the problem statement
precisely.
A new construction of concurrent timestamp systems is presented in
Section~\ref{sec3}, and its correctness proof in Section~\ref{sec4}.
Section~\ref{sec5} concludes the paper.

\section{Model, Problem Definition, and some Notations}\label{sec2}
A concurrent bounded timestamp system (CTS, in short) is
an abstract communication system for $n$ completely asynchronous processes $P_1, \ldots, P_n$.
It consists of $n$ objects ${\cal O}[1..n]$, each of finite space representation, and
supports two operations, namely {\em labeling\/} and {\em
scan(ing)}.
A labeling operation execution (Labeling, in short) of process $P_p$ assigns a new label to
object ${\cal O}[p]$.
It may use all existing labels of ${\cal O}[1..n]$, but it is not allowed to change
the labels of components other than ${\cal O}[p]$.
A scan operation execution (Scan, in short) enables a process to
determine the ordering in which all the objects are written,
that is, it returns a set of labeled-objects ordered temporally\footnote{We ignore, in this paper,
the data values of the objects.}.
It returns a pair $(\overline{l}, \prec)$, where $\overline{l}$ is a set of
current labels, one for each object-component, and $\prec$ is a total order on $\overline{l}$.
Operation executions of each process are sequential.
However, operation executions of different processes need not be sequential, i.e.,
they might overlap.

Let us denote the $k\,$th operation execution 
(Labeling or Scan) of a process $P_{p}$ by $O_{p}^{[k]}$, $k\geq 1$. 
If it is a Scan (Labeling), we denote it explicitly by $S_{p}^{[k]}$ ($L_{p}^{[k]}$).
The label written by a labeling operation execution $L_{p}^{[k]}$ is denoted by $l_{p}^{[k]}$.
 
 For operation executions $A$ and $B$ on a shared
 variable, $A\da B$ means that the execution of $A$ starts before that of $B$
 finishes. That is, if $A\da B$, then either $A\ra B$ or $A$ overlaps $B$;
 in other words, $B\not\ra A$.
 We also assume that if  $B\not\ra A$, then  $A\da B$.
 That is, we assume the global time model \cite{lamp86}.
 
A concurrent timestamp system must ensure the following properties \cite{dole89,gawl92}.
 
\newcounter{bean}
\begin{list}{P\arabic{bean}.}{\usecounter{bean}}
\item Ordering: There exists an irreflexive total order $\Rightarrow$ on the set of all
labeling operation executions, such that the following two conditions hold.
  \begin{itemize}
    \item Precedence: For every pair of Labelings $L_p^{[k]}$ and
    $L_q^{[k']}$, if $L_p^{[k]}\ra L_q^{[k']}$ then $L_p^{[k]}\Rightarrow L_q^{[k']}$.
 
    \item Consistency: For every Scan $S_i^{[j]}$ returning $(\overline{l}, \prec)$, for every two labels  $l_p^{[k]}$ and $l_q^{[k']}$ in $\overline{l}$,
    $l_p^{[k]}\prec l_q^{[k']}$ iff $L_p^{[k]}\Rightarrow L_q^{[k']}$.
    \end{itemize}
 
\item Regularity: For every label $l_p^{[k]}$ in $\overline{l}$ returned by
a Scan $S_i^{[j]}$,
$L_p^{[k]}$ begins before $S_i^{[j]}$ terminates, i.e., $L_p^{[k]}\da
S_i^{[j]}$, and there is no Labeling $L_p^{[k']}$ such that $L_p^{[k]}\ra L_p^{[k']}\ra S_i^{[j]}$.
 
\item Monotonicity: Let $S_i^{[j]}$ and $S_{i'}^{[j']}$ be a pair of Scans
returning sets $\overline{l}$ and $\overline{l'}$, respectively, which contain labels $l_p^{[k]}$
and $l_p^{[k']}$, respectively.
If $S_i^{[j]}\ra S_{i'}^{[j']}$, then $k\leq k'$.
 
\item Extended Regularity: Let $l_p^{[k]}$ be a label returned by a Scan $S_i^{[j]}$.
For each Labeling $L_q^{[k']}$, if $S_i^{[j]}\ra L_q^{[k']}$, then $L_p^{[k]}\Rightarrow L_q^{[k']}$.
\end{list}

The intuitive meaning of the above four properties is as follows.
The ordering property says that all the labeling operation executions can be
totally ordered which is an extension of their real-time precedence order
``$\ra$''.
Moreover, if two different Scans return labels $l$ and $l'$, then both Scans will
have the same order on the labels.
The regularity property says that labels returned by a Scan are not obsolete.
The monotonicity property says that for every two Scans ordered by ``$\ra$'', it is
not the case that the preceding Scan returns a new label of a process $P_p$ and the succeeding
Scan an old label of $P_p$.
The monotonicity property does not imply that labeling and scan operation
executions of all processes are linearizable \cite{herl90}.
It does imply the linearizability of the Scans of all processes and 
labeling operation executions of a single process \cite{dole89}.
The extended regularity property says that if a Scan precedes a labeling
operation execution $L$, then all labels returned by the Scan were assigned by
labeling operation executions that precede $L$ in $\Rightarrow$.

We are interested in those CTSs in which operation executions are {\em wait-free\/}, that is, 
each operation execution will take a bounded number of its own steps (a step is a read/write
of a shared variable), irrespective of the presence of other operation executions and their 
relative speeds.  This paper is concerned with implementing wait-free CTSs from basic 
1-writer 1-reader shared variables.

\section{The Construction} \label{sec3}
For the sake of convenience and better understanding, 
we first present an intuitive informal
description of a construction that uses unbounded shared space \cite{vit86}
(the same idea is used in \cite{dwork92}).
Each process maintains a separate local pool of private values that are natural numbers with
the standard order relations on them.

A label is a vector of $n$ values (`tickets' in \cite{vit86}); its $p$\,th component holds a private
value of process $P_p$.
The current label of ${\cal O}[p]$ is denoted by $l_p[1..n]$ or simply $l_p$.
The current private value of process $P_p$ is $l_p[p]$.
Initially, $l_p[p]=1$ and $l_p[q]=0$, for all $q\not=p$.
To determine a new label for ${\cal O}[p]$, process $P_p$ reads all current private
values of other processes $P_q$, namely $l_q[q]$, and increments its own private value
$l_p[p]$ by one to obtain the new private value.
The new label vector contains these $n$ values, and it is written atomically in
${\cal O}[p]$.
Since the same private value is not used twice in labeling operation
executions, no two labels ever produced in the system are the same.
The ordering
of two label vectors is done by using the standard lexicographic (dictionary) order $\prec$:
for every two labels, $l_p\not=l_q$, the {\em least significant index\/} in which
they differ is the lowest $k$ such that $l_p[k]\not=l_q[k]$;
then, $l_p\prec l_q$ iff $l_p[k]<l_q[k]$.
This lexicographic order $\prec$ is a total order on the set of all possible
labels \cite{fisch}, and this fact is a static common knowledge to the processes.
(In fact, $\prec$ is an elementary example of a well-ordered relation.)
A Scan simply reads all the current labels and orders them using the
lexicographic order.
This unbounded construction satisfies all the properties required for
a concurrent timestamp system (Cf.~\cite{dwork92}).
 
In the unbounded construction discussed above, every time a process  $P_k$
executes a new labeling operation, it uses a new private value
greater than the previously used ones.
In a bounded construction, each process has only a bounded number of private
values, and hence, it needs to use the same
private value at different times, that is, it needs to recycle its own private
values.
The following observation (which is a synthesis of the text
in \cite[page~236]{vit86}) by Dwork and Waarts helps doing the recycling in some possible way.
We quote them verbatim:
\begin{quote}
$\ldots$ for a system to be a concurrent timestamp system, every time a new
private value chosen by process $P_k$ need not be the one that was never used
by $P_k$ beforehand; roughly speaking, instead of increasing its private
value, it is enough for $P_k$ to take as its new private value any value $v$
of its private values that does not appear in any labels, with one proviso:
$P_k$ must inform the other processes that $v$ is to be considered larger than
all its other private values currently in use.
\end{quote}
Consequently, we cannot use the standard ordering relations on the natural numbers
any more, for the numbers may be recycled repeatedly.
One has now to consider these numbers as mere symbols with no standard 
ordering relations defined on them.
We define for every two different private values $v$ and $v'$ of process $P_k$ currently
in use in the system, $v\prec_k v'$ iff $v$ is issued before $v'$ by $P_k$.
Thus, in the bounded construction, the ordering relation 
among the private values
changes in time, and hence it cannot be {\em a priori} common knowledge.
Note that at any point in time, the relation $\prec_k$ on the values
in use is a total order as the values are produced in sequences,
and in fact, it is well-ordered.
For every two labels, $l_p\not=l_q$, obtained by a Scan, if $k$ is  the least
significant index such that $l_p[k]\not=l_q[k]$, then we define $l_p\prec l_q$ iff
$l_p[k]\prec_k l_q[k]$.
Then, $\prec$ is also a well-ordered relation \cite{fisch}.
%Paul: changed from: Then, $\prec$ is also a well-ordered relation \cite{munk87}.
Now, we are concerned with two things in a bounded construction.
First, to make the relations $\prec_k$ useful,
 processes $P_k$ cannot recycle a private value if some other processes are using it.
 Second, for every two private values $v$ and $v'$ of $P_k$ currently in use,
 if $v\prec_k v'$ then all other processes should (get to) know this ordering before using these values.
Note that the meaning of $<$ on the natural numbers is a static common knowledge, but the 
meaning of $\prec_k$ changes continually.
 Thus, every time $P_k$ changes the ordering of two different private values,
 it should inform all the other processes well in advance.
 Then, for all labels read by a Scan, the labels are ordered
 lexicographically, based on the orderings $\prec_k$ of all processes $P_k$.
 Then, the correctness of the bounded system trivially follows from that of the unbounded
 system mentioned above (given in \cite{vit86,dwork92}).

In the following paragraphs, we present a novel construction,
based on \cite{vit86,hald93},
to achieve the afore mentioned two objectives.
The construction is given in Figure~\ref{fig1}.

We now introduce some terminology. 
The description of the construction has five parts: shared variables 
declaration, TRACEABLE-WRITE procedure, TRACEABLE-READ function, 
LABELING procedure and SCAN function.
The procedures and the functions are written in a Pascal-type language. 
To avoid too many `begin's and `end's, some blocks are shown just by indentation.
All the statements in the four routines are numbered only for reference
purposes.

A base shared variable $x$ is read (respectively, written) by executing an instruction
`read {\em local-variable} from $x$' (respectively, `write {\em local-variable} in $x$'),
where the {\em local-variable} is local to the function or the procedure.
The read-instruction assigns the value of $x$ to the {\em local-variable\/},
and the write-instruction writes the value of the {\em local-variable\/} in $x$.
The writer (owner) of a shared variable can retain the value of the variable in its
 local storage and refer to it later on if needed, that is,
it need not read the shared variable to determine the current value of the variable.
Nevertheless, for the sake of convenience and to avoid using many local variables, we let the writer also read its own shared   variable. 
It also uses some private (local, non-shared) variables for each process.
We assume that the private variables are persistent.

Let us consider operation executions of a particular process $P_p$.
Process   $P_{p}$ executes the LABELING procedure to obtain and assign a new label to
${\cal O}[p]$, and executes the SCAN function to report the temporal ordering of
the labels of ${\cal O}[1..n]$.
In a labeling operation execution, it selects a presently unused private value
from its local pool of values (Statements~1--2 in the LABELING procedure),
collects the current  private values of all other processes 
(Statements~5--6), and then writes these
$n$ values atomically in ${\cal O}[p]$ as its new label (Statement~7).
The selection of a new  private value is done in such a way that there is no
trace of this value in the system at present.
In a scan operation execution, process $P_p$ first reads the current labels of all
the processes (Statement~1 in the SCAN function), and then determines their temporal ordering using the latest ordering information
available from some ordering shared variables (Statement~2).

The collection of the current  private values of other processes is done by
executing the TRACEABLE-READ function, and the writing of the new label is done by
executing the TRACEABLE-WRITE procedure.\footnote{These two routines resemble pretty closely
the READ and WRITE routines in \cite{hald95,hald96,vidya90,vit86}.}
These two routines collectively implement atomic reading and writing of labels from and into objects ${\cal O}[p]$.
(In rest of the paper, an execution of the TRACEABLE-READ function (TRACEABLE-WRITE procedure) will be 
called a traceable Read (traceable Write).)
Note that these two routines are not parts of the interface to the CTS, and the processes cannot
directly invoke them.
They directly invoke the LABELING and SCAN routines in which they, in turn, 
invoke traceable 
Read (Write) to read (write) labels.

A process $P_p$ uses shared variables \WC$[p,1..n]$, \RC$[p,1..n]$,
$c[p]$, $label[p,0..1]$ and $copylabel[p,1..n]$ to atomically 
read and write new labels from and into object ${\cal O}[p]$.
The $label$ and $copylabel$ variables are used to hold labels of ${\cal O}[p]$.
\WC\ and \RC\ are handshake variables used to detect overlapping of traceable
Reads and Writes.
The variable $c$ is used to atomically declare writings of new labels
in ${\cal O}[p]$.
Process $P_p$ uses the shared variables $order[p,1..n]$ to inform all the processes
of the latest ordering relation $\prec_p$.
The shared variables $lend[p,1..n]$ are used to inform all the processes
which of their private values might be in use in the system.
The component $lend[p,j]$ contains all the private values of process $P_j$
that $P_p$ may have lent to other processes.
Process $P_p$ also uses static private variables: $cl_p$, $myLend_p$,
$\prec_p$, and $old$-$label_p$. 
$cl_p$ and $myLend_p$ always store the values of $c[p]$ and $lend[p,1..n]$,
respectively, locally.
$\prec_p$ contains the latest ordering information of all the private values in use in the system.
$old$-$label_p$ stores the label of the on-going or the recently completed
Labeling operation execution.

The traceable Writes of process $P_p$
use two $n$-reader safe {\em main label variables\/}, 
$label[p,0]$ and $label[p,1]$, and 
a 1-reader safe {\em copy label variable\/} for each process, $copylabel[p,1..n]$. 
% These variables along with a multireader boolean atomic variable $c[p]$
% are used to implement atomic Writes of new labels by $P_p$
% in object ${\cal O}[p]$ (Cf.~\cite{vidya90,hald95,hald96}).
The main label variables are used alternately for writing successive new labels.
 Immediately after writing a
new label in a main label variable, the process records that variable index
in the 1-writer multireader boolean atomic variable $c[p]$. 
(This writing atomically `declares' the current label of component ${\cal O}[p]$.)
Then the process checks for each $i$ whether a new traceable Read of
process $P_{i}$ started since the last traceable Write (of $P_p$). This is done by using a pair of
boolean 1-writer 1-reader (handshaking) atomic variables \RC$[i,p]$ and \WC$[p,i]$.\footnote{This 
strategy of detecting overlapping operation execution is pioneered by
Peterson \cite{pete83}.} 
Process $P_i$ sets these
values different, by assigning the complement of \WC$[p,i]$ to \RC$[i,p]$,
at the beginning of each traceable Read (Statements~1--2 in TRACEABLE-READ),
and process $P_p$ makes sure that they are the same, at the end of each traceable Write (Statements~4.1 and 4.2.3 in TRACEABLE-WRITE).
By this way the processes $P_p$ and $P_i$ can find if there are overlappings of
their traceable Writes and Reads.
Hence if the two values are different when the process $P_p$ checks them, a new traceable Read of $P_{i}$ must have started by then.
In that case, $P_p$ writes the new label value in $copylabel[p,i]$ also, 
and then sets the above values the same,
by assigning the \RC$[i,p]$ value to \WC$[p,i]$. 
(This way it is guaranteed that a reading and a writing on $copylabel$ variables do not 
overlap each other, and contains a valid value for the traceable Read \cite{vit86,hald95,vidya90}.)
For each such process $P_{i}$, $P_p$ takes a note which of the private values of
processes $P_j$ could be used by $P_i$ (Statement~4.2.2).
Finally, $P_p$ informs all the processes $P_j$ which of their private values could
be in use (all that $P_p$ knows of) through 1-writer 1-reader regular variables $lend[p,j]$ (Statement~6).

Each traceable Read of process $P_{p}$, from a process
$P_i$, after reading \WC$[i,p]$ and writing its
complement in \RC$[p,i]$ as mentioned above (Statements~1--2 in TRACEABLE-READ), finds out from $c[i]$ the main 
label variable
that has been written by $P_i$ most recently, and reads from that variable.
Then it reads \WC$[i,p]$ again and compares with \RC$[p,i]$.
If the two values continue to be different, then the reading of the main label variable
does not overlap any writings of the label variable and hence it returns the value
just read from the main label variable. 
Otherwise, there is a possibility that the reading of the label variable overlaps with some
writing of the same variable, and hence, it reads $copylabel[i,p]$ and returns that value.
Note that in the latter case, a traceable Write by $P_i$ must have finished 
(with respect to $P_p$, that is, $P_i$ must have done loop iteration $p$ at
Statement~4 in TRACEABLE-WRITE) after the traceable Read started, and 
that Write would have written in $copylabel[i,p]$.

In selecting a new (currently unused) private value, process $P_p$ does not use any of  the values
%Paul: Changed "all" to "any": In selecting a new (currently unused) private value, process $P_p$ does not use all the values
stored in $lend[1..n,p]$ (Statements~1--2 in LABELING).
After selecting the  new private value, say $v$, $P_p$ informs all processes
$P_i$ that $v$ is the most recent private value through 1-writer 1-reader
regular variables $order[p,i]$ (Statements~3--4)
which are used by the Scans of $P_i$.

\section{Correctness Proof} \label{sec4}
\begin{proposition}
\mbox{\bf \cite{lamp86}} For operation executions $B$ and $C$ on a
shared variable, and all operation executions $A$ and $D$, if 
$A\ra B\da C\ra D$, then $A\ra D$.
\end{proposition}
\noindent{\em Proof:\/} The implication follows by the transitivity of (i)~$A$
finishes before $B$ starts, (ii)~$B$ starts before $C$ finishes and (iii)~$C$
finishes before $D$ starts. $\Box$

\noindent{\bf Definition.}
For operation executions $A$ and $B$ executed on the same atomic variable $x$,
we say
$A \RA{x} B$ if $A$ precedes $B$ in the total ordering imposed on the
operation executions by the atomic variable.
The subscript $x$ is omitted when it is clear from the context. $\Box$

\begin{proposition}
For operation executions $B$ and $C$ on an atomic variable $x$, and all
operation executions $A$ and $D$, if $A\ra B\RA{x} C\ra D$, then $A\ra D$.
\end{proposition}
\noindent{\em Proof:\/}
The relation $B\RA{x}C$ implies $B$ precedes or overlaps $C$ (since the total
order imposed on the operation executions by the atomic variable
is an extension of the
precedence relation), that is, $B\da C$.
Then the implication follows by Proposition~1. $\Box$

The following notations are used in the presentation of the correctness
proofs.
\setcounter{bean}{1}
\begin{list}{N\arabic{bean}.}{\usecounter{bean}}
\item The {\em k\,}th operation execution of a process $P_p$ is denoted, as
stated in Section~\ref{sec2}, by $O_p^{[k]}({\cal O})$, $k\geq 1$; if it is a Scan
(alternatively, a Labeling), we denote it explicitly by $S_{p}^{[k]}({\cal O})$
(alternatively,
$L_{p}^{[k]}({\cal O})$).
The `$({\cal O})$' part in the notation is omitted when it is clear from the context.
All the operation executions of $P_p$ are totally ordered.
That is, for $k>2$, $O_p^{[k-1]}\ra O_p^{[k]}$.
% That is, for $k>1$, $O_p^{[k-1]}\ra O_p^{[k]}$.
% (To avoid ambiguities, we assume the existence of a fictitious operation
% execution $O_p^{[0]}=L_{p}^{[0]}$ that writes the initial label.
% The operation execution $L_{p}^{[0]}$ takes place before any non fictitious
% operation execution starts.)

\item For a shared variable $x$, the
Read (respectively, Write) of $x$ by $O_p^{[k]}$ is denoted by $R_p^{[k]}(x)$
(respectively, $W_p^{[k]}(x)$).
If $x$ is referred more than once, then the superscript $[k,j]$ is used for
the
{\em j\,}th access.

\item Each operation execution $O_{p}^{[k]}$ ($L_{p}^{[k]}$ or $S_{p}^{[k]}$)
of process $P_p$ executes the TRACEABLE-READ function for every other process
$P_i$; the whole function execution is denoted by a traceable Read
$TR_{p,i}^{[k]}$.

\item Each labeling operation execution  $L_{p}^{[k]}$ of process $P_p$
executes the TRACEABLE-WRITE procedure; the whole procedure execution is
denoted by a traceable Write $TW_{p}^{[k]}$.

\item For the sake of convenience, the variables \RC$[p,i]$ and \WC$[p,i]$ are
abbreviated to $r_{p,i}$  and $w_{p,i}$, respectively.
\end{list}

\noindent{\bf Definition.}
For a shared variable $x$, we define a {\em reading mapping\/} $\pi_x$ for
Reads of $x$ as follows: if a Read $R$ returns the value written by a Write
$W$, then $\pi_x(R)$ is $W$; otherwise $\pi_x(R)$ is undefined.
(Note, for safe $x$, $\pi_x$ is a partial mapping.)
We omit the subscript $x$ when it is clear from the context. 
$\Box$

\begin{lemma} \label{lem1}
(a) No two consecutive labeling operation executions of a process have the 
same  private value. 

(b) No two consecutive traceable Writes of
a process have the same private value. 
\end{lemma}
\noindent{\em Proof:\/} Part~($a$) follows from the select statement (Statement~2) in the LABELING procedure.
Part~($b$) follows from Part~($a$) as each Labeling executes 
one and the only one traceable Write.~$\Box$

\begin{lemma}\label{lem2}
Each time the value written in \wpi\ is the complement of the previous value of \wpi.
\end{lemma}
\noindent{\em Proof:\/}
Immediate from Statements~4.1, 4.2 and 4.2.3 in
the TRACEABLE-WRITE procedure. $\Box$

\begin{lemma}\label{lem3}
Any traceable Write $TW_{p}^{[k]}$ (actually, $L_{p}^{[k]}$) that writes \wpi\ sets $\wpi =\rpi$, and
if $R_{i}^{[l,1]}(\wpi)\LA W_{p}^{[k]}(\wpi)\LA R_{i}^{[l,2]}(\wpi)$
for some traceable Read $TR_{i,p}^{[l]}$ (actually, $O_i^{[l]}$) of process $P_{i}$, then the equality continues to hold until
the execution of $TR_{i,p}^{[l]}$ is complete, in fact until the next traceable Read $TR_{i,p}^{[l+1]}$ writes \rpi.
\end{lemma}
\noindent{\em Proof:\/} 
Initially, $\wpi=\rpi$, since both of them  are initialized to 0. 
Among the traceable Writes of the process $P_{p}$, some will write \wpi, and some will not.
Let $TW_{p}^{[k_{j}]}$, $j\geq 1$, $k_{j}\geq 1$, be the $j$\,th traceable Write that writes \wpi.

Consider $TW_{p}^{[k_{1}]}$. 
By Lemma~\ref{lem2}, it writes 1 in \wpi. 
This implies, by Statements~4.1 and 4.2.3 in TRACEABLE-WRITE, that it read 1 from \rpi. Since the initial value of \rpi\ is
0, some traceable Read of $P_{i}$ must have written 1 in \rpi. 
Let $TR_{i,p}^{[l_{1}]}$ be the first such traceable Read. Then $W_{i}^{[l_{1}]}(\rpi)\LA R_{p}^{[k_{1}]}(\rpi)$.
Note that $TR_{i,p}^{[l_{1}]}$ reads 0 from \wpi\ and hence writes 1 in \rpi\
(Statements~1--2 in TRACEABLE-READ).
Also each subsequent traceable Read $TR_{i,p}^{[l'_{1}]}$, if any, such that
$R_{i}^{[l'_{1},1]}(\wpi)\LA W_{p}^{[k_{1}]}(\wpi)$, would read 0 from \wpi, and hence
will write 1 in \rpi. 
Hence, irrespective of whether $W_{i}^{[l'_{1}]}(\rpi)\LA R_{p}^{[k_{1}]}(\rpi)$ or
$R_{p}^{[k_{1}]}(\rpi)\LA W_{i}^{[l'_{1}]}(\rpi)$,
on $W_{p}^{[k_{1}]}(\wpi)$, 
$\wpi=\rpi$, and if
$R_{i}^{[l,1]}(\wpi)\LA W_{p}^{[k_{1}]}(\wpi)\LA R_{i}^{[l,2]}(\wpi)$ for some traceable Read $TR_{i,p}^{[l]}$,
then the equality continues to hold until  $TR_{i,p}^{[l]}$ is complete, in fact until 
the next traceable Read $TR_{i,p}^{[l+1]}$ writes \rpi,
since \wpi\ will not be changed by any traceable Write $TW_{p}^{[k_{1}']}$, for $k_{1}'>k_{1}$,
that may occur before  $TR_{i,p}^{[l]}$ is complete.

% Assuming as induction hypothesis that the assertion holds for
% $TW_{p}^{[k_{j}]}$, for some $j$, we can show in a similar fashion that the assertion holds
% for $TW_{p}^{[k_{j+1}]}$.  $\Box$

Assuming as induction hypothesis that the assertion holds for
$TW_{p}^{[k_{j}]}$, for some $j$, we show that the assertion holds
for $TW_{p}^{[k_{j+1}]}$.
By the statement of the lemma, $TW_{p}^{[k_{j}]}$ sets $\wpi=\rpi$
by writing value, say $b\in\{0,1\}$ in $\wpi$.
Then, by Lemma~\ref{lem2}, $TW_{p}^{[k_{j+1}]}$ writes $\neg b$ 
in $\wpi$.\footnote{$\neg b$ is defined as $1-b$.}
This implies by Statements~4.1 and 4.2.3 in TRACEABLE-WRITE, it read $\neg b$
from $\rpi$.
As the value of $\rpi$ is $b$ when $TW_{p}^{[k_{j}]}$ reads it,
there must be a traceable Read that writes $\neg b$ in $\rpi$
after $TW_{p}^{[k_{j}]}$ sets $\wpi=\rpi$.
Let $TR_{i,p}^{[l]}$ be the first such traceable Read.
Then, $W_{i}^{[l]}(\rpi)\LA R_{p}^{[k_{j+1}]}(\rpi)$, and
$TR_{i,p}^{[l]}$ writes $\neg b$ in $\rpi$.
Each subsequent traceable Read $TR_{i,p}^{[l']}$, if any, such that
$R_{i}^{[l',1]}(\wpi)\LA W_{p}^{[k_{j+1}]}(\wpi)$, would read $b$ from \wpi, and hence
will write $\neg b$ in \rpi.
Hence, irrespective of whether $W_{i}^{[l']}(\rpi)\LA R_{p}^{[k_{j+1}]}(\rpi)$ or
$R_{p}^{[k_{j+1}]}(\rpi)\LA W_{i}^{[l']}(\rpi)$,
on $W_{p}^{[k_{j+1}]}(\wpi)$,
$\wpi=\rpi$. 
If $R_{i}^{[l]}(\wpi)\LA W_{p}^{[k_{j+1}]}(\wpi)\LA R_{i}^{[l,2]}(\wpi)$ for some traceable Read $TR_{i,p}^{[l]}$,
then the equality continues to hold until  $TR_{i,p}^{[l]}$ is complete, in fact until
the next traceable Read $TR_{i,p}^{[l+1]}$ writes \rpi,
since \wpi\ will not be changed by any traceable Write $TW_{p}^{[k']}$, for $k'>k_{j+1}$,
that may occur before  $TR_{i,p}^{[l]}$ is complete.
$\Box$

Lemma~\ref{lem3} implies the following property.

\begin{lemma}\label{lem4}
Let $TR_{i,p}^{[l]}$ be a traceable Read. 
There can be at most one traceable Write, say $TW_{p}^{[k]}$, such that
$R_{i}^{[l,1]}(\wpi) \LA W_{p}^{[k]}(\wpi)\LA R_{i}^{[l,2]}(\wpi)$. 
The traceable Read $TR_{i,p}^{[l]}$ on $R_{i}^{[l,2]}(\wpi)$ will find
$\rpi=\wpi$ if there is such a traceable Write, and $\rpi\not=\wpi$ otherwise. $\Box$ 
\end{lemma}

In the following we use a typical kind of notation for labeling operation
executions.
\begin{list}{N\arabic{bean}.}{\usecounter{bean}}
\setcounter{bean}{5}
\item The labeling operation executions of  process $P_p$ are sometimes
denoted by $L_p^{[k_j]}$, where $k$ is some alphabet and $j$ is a natural
number, $j\geq 1$, $k_j\geq 1$.
Thus, for $j>1$, $L_p^{[k_{j-1}]}$ and $L_p^{[k_j]}$ are two consecutive
labeling operation executions of $P_p$ such that $L_p^{[k_{j-1}]}\ra
L_p^{[k_j]}$. They need not be two consecutive
operation executions, that is, $k_{j}\geq k_{j-1}+1$.
\end{list}

In the following two lemmas, we show that traceable Reads return valid
label values. We also define their reading mapping function $\pi$. Lemmas~\ref{lem5} and \ref{lem6}
deal with the case traceable Reads return values from $label$ and $copylabel$ variables, respectively.

\begin{lemma}\label{lem5}
Let $TR_{i,p}^{[l]}$ be a traceable Read that finds $\rpi\not=\wpi$ on $R_{i}^{[l,2]}(\wpi)$. 
Suppose $\pi(R_{i}^{[l]}(c[p]))$ is $W_p^{[k_j]}(c[p])$ (of the traceable
Write $TW_p^{[k_j]}$ of $L_p^{[k_j]}$), and  $label[p,x]$ is the main label variable from which
$TR_{i,p}^{[l]}$ returns the label value.
\newcounter{cntr1} 
\begin{list}{(\alph{cntr1})}{\usecounter{cntr1}} 
\item If $j'$ is the least index such that $R_{i}^{[l,2]}(\wpi)\LA W_{p}^{[k_{j'}]}(\wpi)$, then $j'$ equals $j$ or $j+1$.

\item $\pi(TR_{i,p}^{[l]})$ is $TW_p^{[k_j]}$.

\item The traceable Read $TR_{i,p}^{[l]}$ reading $label[p,x]$ does not conflict with any 
traceable Write writing that label variable.
\end{list}
\end{lemma}

\noindent{\em Proof:\/}  

\noindent{($a$)} 
Let $j''$ be the greatest index such that $j''<j'$ and $TW_{p}^{[k_{j''}]}$ writes \wpi.
Then by (i)~the choice of $j'$, 
(ii)~the assumption that $TR_{i,p}^{[l]}$ finds $\rpi\not=\wpi$ on $R_{i}^{[l,2]}(\wpi)$
and 
(iii)~Lemma~\ref{lem4}, 
it follows that 
$W_{p}^{[k_{j''}]}(\wpi)\LA R_{i}^{[l,1]}(\wpi)$. That is,
$W_{p}^{[k_{j''}]}(\wpi)\LA R_{i}^{[l,1]}(\wpi)\ra R_{i}^{[l,2]}(\wpi)\LA W_{p}^{[k_{j'}]}(\wpi)$.
The traceable Write $TW_{p}^{[k_{j''}]}$ sets \wpi\ equal to \rpi, 
$TR_{i,p}^{[l]}$ sets \rpi\ not equal to \wpi,  and
hence $TW_{p}^{[k_{j'}]}$ is the first traceable Write, after $TW_{p}^{[k_{j''}]}$, that finds
$\rpi\not=\wpi$.

 From $W_{i}^{[l]}(\rpi)\ra R_{i}^{[l]}(c[p])\LA W_p^{[k_{j+1}]}(c[p])\ra R_{p}^{[k_{j+1}]}(\rpi)$, we
have $W_{i}^{[l]}(\rpi)\ra R_{p}^{[k_{j+1}]}(\rpi)$.
That is, the traceable Write $TW_p^{[k_{j+1}]}$ will find $\rpi\not=\wpi$, the inequality set by $TR_{i,p}^{[l]}$,
unless an earlier traceable Write has found the inequality and set \wpi\ equal to \rpi.
We claim that such an earlier traceable Write, if one exists, can only be $TW_p^{[k_j]}$. 
Suppose, on the contrary, that it is $TW_{p}^{[k_{j'''}]}$, for $j'''<j$.
Then, 
by the choice of $j''$ and Lemma~\ref{lem4}, we have
$W_{p}^{[k_{j''}]}(\wpi)\LA R_{i}^{[l,1]}(\wpi)\ra R_{i}^{[l]}(c[p])\ra R_{i}^{[l,2]}(\wpi)\LA W_{p}^{[k_{j'''}]}(\wpi)\ra W_p^{[k_j]}(c[p])$.
This implies $R_{i}^{[l]}(c[p])\ra W_{p}^{[k_j]}(c[p])$, contradicting the assumption that $\pi(R_{i}^{[l]}(c[p]))$ is $W_p^{[k_j]}(c[p])$.  
The assertion follows.

\noindent{($b$ and $c$)} 
Let $label[p,x']$ be the variable in which $TW_p^{[k_j]}$ writes.

For $j'$ described in part~($a$), we have
$R_{i}^{[l]}(label[p,x])\ra R_{i}^{[l,2]}(\wpi)\LA W_{p}^{[k_{j'}]}(\wpi)\ra TW_{p}^{[k_{j+2}]}$.
That is, $TR_{i,p}^{[l]}$  finishes reading $label[p,x]$ before the
traceable Write $TW_{p}^{[k_{j+2}]}$ starts its execution.
 From (i)~the assumption that $\pi(R_{i}^{[l]}(c[p]))$ is $W_{p}^{[k_{j}]}(c[p])$, (ii)~the property that
$TW_{p}^{[k_{j+1}]}$ does not write in the same main label variable that $TW_p^{[k_j]}$ writes,
\linebreak (iii)~$W_p^{[k_j]}(label[p,x'])\ra W_p^{[k_j]}(c[p])\LA R_{i}^{[l]}(c[p])\ra
R_{i}^{[l]}(label[p,x])$,
and (iv)~Statements~1--3 in TRACEABLE-WRITE,
it follows that $x=x'$, and $TW_p^{[k_j]}$ finishes writing $label[p,x]$ before $TR_{i,p}^{[l]}$ starts reading it. 
The assertions follow.
$\Box$

\begin{lemma}\label{lem6}
Let $TR_{i,p}^{[l]}$  be a traceable Read that finds $\rpi=\wpi$ on $R_{i}^{[l,2]}(\wpi)$. 
Suppose $TW_{p}^{[k_{j}]}$ is the traceable Write such that
$R_{i}^{[l,1]}(\wpi)\LA W_{p}^{[k_{j}]}(\wpi)\LA R_{i}^{[l,2]}(\wpi)$. 
\newcounter{cntr2} 
\begin{list}{(\alph{cntr2})}{\usecounter{cntr2}} 
\item The traceable Read  $TR_{i,p}^{[l]}$ reading $copylabel[p,i]$  does not conflict with any  traceable Write writing it.

\item $\pi(TR_{i,p}^{[l]})=TW_{p}^{[k_{j}]}$. 
\end{list}
\end{lemma}

\noindent{\em Proof:\/} ($a$ and $b$)
By Lemma~\ref{lem4}, $TW_{p}^{[k_{j}]}$ is the only traceable Write such that $R_{i}^{[l,1]}(\wpi)\LA W_{p}^{[k_{j}]}(\wpi)\LA R_{i}^{[l,2]}(\wpi)$. 
It is clear from the TRACEABLE-WRITE procedure that $TW_{p}^{[k_{j}]}$ writes the
value in $copylabel[p,i]$ (Statement~4.2.1) before setting the $\wpi$ and $\rpi$ values equal (Statement~4.2.3).
This equality will not be changed until $P_i$ starts the next traceable Read.
Thus, the traceable Write $TW_{p}^{[k_{j+1}]}$ and subsequent traceable Writes of $P_{p}$,
if they find $\rpi=\wpi$, will not write the copy label variable.
 From $W_{p}^{[k_{j}]}(copylabel[p,i])\ra W_{p}^{[k_{j}]}(\wpi)\LA R_{i}^{[l,2]}(\wpi)\ra R_{i}^{[l]}(copylabel[p,i])$, we have
$W_{p}^{[k_{j}]}(copylabel[p,i])\ra R_{i}^{[l]}(copylabel[p,i])$.
The assertions follow.
$\Box$

Now we would like to show that private values of processes $P_p$ are traceable.
If a process $P_i$ in its current label uses a private value $v$ of another
process $P_p$, $P_i$ informs this ``using of'' $v$ by setting $lend[i,p][1][i]$ to
$v$ at the end of the corresponding traceable Write (Statements~5--6).
Thus, all the private values in the existing labels are traceable
by their respective owners.
The following lemma shows that the  private values used by Scans are also
traceable.

\begin{lemma}\label{lem7}
Let a Scan $S_i^{[l]}$ of a process $P_i$
use a private value $v$ of a process $P_p$ that has written
the value $v$ in a traceable Write $TW_p^{[k_j]}$. Then,
$P_p$ does not recycle $v$ until $S_i^{[l]}$ is complete.
\end{lemma}
\noindent{\em Proof:\/}
We need to consider the following two cases.

\noindent{\em Case~1:} $S_i^{[l]}$ got $v$ directly from $P_p$. 

We need to consider two subcases.

\noindent{\em Subcase a.}
If the traceable Read $TR_{i,p}^{[l]}$ returns the value $v$ from
$copylabel[p,i]$,
then, by Lemma~\ref{lem6} and \ref{lem4}, the traceable Write $TW_p^{[k_j]}$ has executed the {\em if}-statement body (Statement~4.2) for process $P_i$.
There it has set $myLend_p[p][1][i]$ to $v$ (Statement~4.2.2).
The successive traceable  Writes of $P_p$ that occur before $S_i^{[l]}$ is
complete will not execute the if-statement, and hence, will not change the $myLend_p[p][1][i]$ value.
(Statement~5 does not change the value too.)
As the labeling operation executions of $P_p$ do not reuse the values
referred to in $lend[1..n,p]$, $v$ will not be reissued at least until $S_i^{[l]}$
is complete (Statements~1--2 in LABELING).

\noindent{\em Subcase b.}
If the traceable Read $TR_{i,p}^{[l]}$ returns the value $v$ from a main label
variable, then by Lemma~\ref{lem5}($a$), traceable Write $TW_p^{[k_j]}$ or  $TW_p^{[k_{j+1}]}$ executes the
{\em if}-statement for process $P_i$.
In the case of $TW_p^{[k_j]}$, $myLend_p[p][1][i]$ is set to $v$, and in the case of
$TW_p^{[k_{j+1}]}$, $myLend_p[p][0][i]$ is set to $v$ (Statements~4.2.2 and 7). 
The successive traceable  Writes of $P_p$ that occur before $S_i^{[l]}$ is
complete will not execute the if-statement, and hence, will not change the $myLend_p[p][0..1][i]$ values.
(Statement~5 does not change the values too.)
By Lemma~\ref{lem1}, $TW_p^{[k_{j+1}]}$ uses a private value different from $v$.
So, by the argument given in the
Subcase~$a$, $v$ will not be reissued as a new private
value until $S_i^{[l]}$ is complete.

\noindent{\em Case~2:} $S_i^{[l]}$ got $v$ from another process $P_q$.

\noindent{\em Claim}. Process $P_q$ has obtained $v$ directly from $P_p$.\\
\noindent{\em Proof}:
Note $S_i^{[l]}$ got $v$ by reading a label from $P_q$.
That is, $P_q$ writes $v$ in the $p\,th$ component of the label.
To form a new label, $P_q$ uses the $j$\,th component of the labels
it reads from processes $P_j$ (Statements~5--6 in LABELING).
Hence, $P_q$ obtains $v$ directly from $P_p$.
$\Box$

Let $L_q^{[m_o]}$ be the corresponding labeling operation execution.
Note that each labeling operation execution also executes traceable Reads
(Statement~5).
Then $\pi(TR_{q,p}^{[m_o]})$ is $TW_p^{[k_j]}$ and $\pi(TR_{i,q}^{[l]})$ is $TW_q^{[m_o]}$.
As argued in Case~1, either $TW_p^{[k_j]}$ or $TW_p^{[k_{j+1}]}$ stores $v$ in $myLend_p[p][0..1][q]$.
This value will not be changed until $L_q^{[m_o]}$ is complete, in fact until $P_q$
starts its next operation execution $O_q^{[m_o+1]}$.
Let $TW_p^{[k_{j'}]}$, $j'\geq j+1$,  be the first traceable Write that changes the $myLend_p[p][0..1][q]$ values different from $v$.
Then, it must have found $L_q^{[m_o]}$ is complete and the next operation
execution of $P_q$, namely $O_q^{[m_o+1]}$, has started.
 From $W_q^{[m_o]}(lend[q,p])\ra O_q^{[m_o+1]}({\cal O})\da L_p^{[k_{j'}]}({\cal O})\ra L_p^{[k_{j'+1}]}$,
we have $W_q^{[m_o]}(lend[q,p])\ra L_p^{[k_{j'+1}]}$.
That is, $L_p^{[k_{j'+1}]}$ and successive labeling operation executions of $P_p$
would not reissue $v$ if $v$ is found in $lend[q,p]$ (Statements~1--2).
Note that $TW_q^{[m_o]}$ will write $v$ in $lend[q,p][1][q]$ at the end of its execution (Statements~5--6 in TRACEABLE-WRITE).
Also note that the traceable Write $TW_p^{[k_{j'}]}$ (actually $L_p^{[k_{j'}]}$) does not issue $v$.
Now, from $\pi(TR_{i,q}^{[l]})$ is $TW_q^{[m_o]}$ it follows, by
Lemmas~\ref{lem5} and \ref{lem6}, that either $TW_q^{[m_o]}$
or $TW_q^{[m_{o+1}]}$ would execute the {\em if}-statement for $P_i$, and write $v$ in $myLend_q[p][0..1][i]$ indicating that the private value $v$ of $P_p$
is being used by $P_i$, 
and this will not be
changed until $S_i^{[l]}$ is complete; in fact, until the next operation 
execution $O_{i}^{[l+1]}$ of $P_i$ starts.
Hence $L_p^{[k_{j'+1}]}$ and successive labeling operation executions of $P_p$
that may occur before $S_i^{[l]}$ is complete are able to trace $v$ in $lend[q,p]$,
and hence, will not reissue $v$.
$\Box$

\begin{corollary}\label{cor1}
It is clear from the proof of Lemma~\ref{lem7} that if a Scan $S_i^{[l]}$ uses
a private value $v$ of $P_p$ which is written in labeling operation execution
 $L_p^{[k_{j}]}$, then $TW_p^{[k_j]}({\cal O}[p])\da TR_{i,p}^{[l]}({\cal O}[p])$ for direct reading and
$TW_p^{[k_j]}({\cal O}[p])\da TR_{q,p}^{[m_o]}({\cal O}[p])\ra TW_q^{[m_o]}({\cal O}[q])\da TR_{i,q}^{[l]}({\cal O}[q])$
for indirect reading of $v$ via process $P_q$.
For the latter relation, by the axioms of Anger\cite{ange89}, 
$TW_p^{[k_j]}({\cal O}[p])\da TR_{i,q}^{[l]}({\cal O}[q])$
$\Box$
\end{corollary}

The following lemma shows that Scans can determine the correct temporal order of the
private values of all processes.

\begin{lemma}\label{lem8}
Let  $S_i^{[l]}$ be a Scan that  uses private values $v$ and $v'$ of
a process $P_p$.
Then, $S_i^{[l]}$ can determine the correct temporal order between the values $v$ and $v'$.
\end{lemma}
\noindent{\em Proof:\/}
Assume Scan $S_i^{[l]}$ uses the two different private values $v$ and $v'$ of 
process $P_p$ that has written them in traceable Writes $TW_p^{[k_j]}$ and
$TW_p^{[k_{j'}]}$, respectively, where $j<j'$, and hence, $v\prec_p v'$ (as
defined in Section~\ref{sec3}).
By Lemma~\ref{lem7}, $P_p$ does not recycle $v$ and $v'$ until $S_i^{[l]}$ is
complete.
To guarantee the correctness of the timestamp system, we need to make sure
that $S_i^{[l]}$ can correctly determine the order $v\prec_p v'$ in case these values are used in ordering some of the
scanned labels.
 From the LABELING and SCAN routines and Corollary~\ref{cor1}, we have $W_p^{[k_{j'}]}(order[p,i])\ra TW_p^{[k_{j'}]}({\cal O}[p])\da
TR_{i,q}^{[l]}({\cal O}[q])\ra R_{i}^{[l]}(order[p,i])$, where $q$ is 
as defined in Corollary~\ref{cor1}.
That is,
$W_p^{[k_{j'}]}(order[p,i])\ra R_{i}^{[l]}(order[p,i])$.
Now, we need to make sure that $L_p^{[k_{j'}]}$ can correctly determine
that the private value $v$ is being used by the process $P_i$, before writing $order[p,i]$.
Off course, it would assume $v'$ could be used by $P_i$ too.
As it knows $v\prec_p v'$, to inform this ordering to $P_i$, it writes $v$ at
a lower indexed entry in $order[p,i]$ than $v'$. 
The successive labeling operation executions do not change this ordering.
Thus, $P_i$ can determine the order of $v$ and $v'$ correctly after reading 
$order[p,i]$, by the regularity of order variables.

Now we answer the question how  $L_p^{[k_{j'}]}$ finds that $v$ might be
used by $P_i$.
Note that $P_p$ does not know precisely which of its private values $P_i$ is
going to use.
So, it guesses a subset of its private values, which contains the values actually being
used by $P_i$.
There are two cases to be considered.

\noindent Case~1. $P_i$ obtains $v$ directly from  $P_p$. Either $TW_p^{[k_{j}]}$ or
$TW_p^{[k_{j+1}]}$ will reserve $v$ for $P_i$ by storing $v$ in $lend[p,p][0..1][i]$,
and hence the use of $v$ by $P_i$ is traceable.

\noindent Case~2. $P_i$ obtains $v$ indirectly through another process $P_q$, for some $q$.
 From the claim in the proof of Lemma~\ref{lem7}, we know that $P_q$ has 
obtained $v$ directly from  $P_p$.
Let the corresponding labeling operation execution be $L_q^{[m_0]}$.
Either  $TW_p^{[k_{j}]}$ or
$TW_p^{[k_{j+1}]}$ will set $lend[p,p][0..1][q]$ to $v$, and $P_p$ assumes $v$
could be used by any process $P_i$ through ${\cal O}[q]$
(one level of indirect propagation of a private value).
At the end of $L_q^{[m_o]}$, in $TW_q^{[m_o]}$, $P_q$ informs $P_p$ that $v$ is in ${\cal O}[q]$ by setting
$lend[q,p][1][q]$ to $v$ (Statements~5--6), and this value could be used by any process $P_i$.
Alternatively, if $P_q$ detects that the $v$ is being used by $P_i$, it informs
 ``this using'' through $lend[q,p][0..1][i]$ (Statements~4.2.2 and 6).

Hence, if $L_p^{[k_{j'}]}$ finds $v$ in  $lend[p,p][0..1][i]$ or
 $lend[p,p][0..1][q]$ or $lend[q,p][1][q]$ or $lend[q,p][0..1][i]$, 
for some $q$, it will assume that $v$ is being used by $P_i$
(Statements~1 and 4.1 in LABELING procedure).

The assertion follows.~$\Box$

\begin{claim}\label{claim1}
Each order variable is of size at most $5n$.
\end{claim}
\noindent{\em Proof\/}:
As discussed in the proof of Lemma~\ref{lem8}, $P_p$ needs to reserve
its private values referred to in $lend[q,p][0..1][i]$, $lend[q,p][1][q]$ and
$lend[p,p][0..1][q]$ for all $q$, that is, at most $5n$ values for process $P_i$.
The claim follows.
$\Box$

\begin{corollary}\label{cor2}
The set of private values is bounded.
In fact, by Statements~1--2 in the LABELING procedure, 
the size of the set is less than $2n^2$.
$\Box$
\end{corollary}

By the discussion at the end of 3\,rd paragraph, Section~\ref{sec3}, the correctness of the
proposed construction is immediate. However, for the sake of completeness, we give the proof in Theorem~\ref{theo1}.
Before that a technical lemma follows.

\begin{lemma}\label{lem9}
Let $TR_{i,p}^{[l]}$ and $TR_{i',p}^{[l']}$ be two traceable Reads such that
$TR_{i,p}^{[l]}\ra TR_{i',p}^{[l']}$ and $\pi(TR_{i,p}^{[l]})$ be
$TW_{p}^{[k_j]}$. Then,
\begin{list}{(\alph{cntr2})}{\usecounter{cntr2}}
\item $W_{p}^{[k_j]}(c[p])\LA R_{i'}^{[l']}(c[p])$,
\item $\pi(TR_{i',p}^{[l']})$ is $TW_{p}^{[k_{j'}]}$, where $j'\geq j$,
$k_{j'}\geq k_j$.
\end{list}
\end{lemma}
\noindent{\em Proof\/}:
We have the following two cases.

\noindent{\em Case~1:\/} $TR_{i,p}^{[l]}$ finds $\rpi\not=\wpi$ on
$R_{i}^{[l,2]}(\wpi)$.

Lemma~\ref{lem5}($b$) implies that $\pi(R_{i}^{[l]}(c[p]))$ is
$W_{p}^{[k_j]}(c[p])$.
Then, we have $TW_{p}^{[k_{j-1}]}\ra W_{p}^{[k_j]}(c[p])\LA 
R_{i}^{[l]}(c[p])\ra R_{i'}^{[l',1]}(w_{p,i'})\ra R_{i'}^{[l']}(c[p])$.

\noindent{\em Case~2:\/} $TR_{i,p}^{[l]}$ finds $\rpi=\wpi$ on $R_{i}^{[l,2]}(\wpi)$.

By Lemma~\ref{lem6}, we have 
$TW_{p}^{[k_{j-1}]}\ra W_{p}^{[k_j]}(c[p])\ra W_{p}^{[k_j]}(\wpi)\LA 
R_{i}^{[l,2]}(\wpi)\ra R_{i'}^{[l',1]}(w_{p,i'})\ra R_{i',p}^{[l']}(c[p])$. 

For both the cases we have $W_{p}^{[k_j]}(c[p])\LA R_{i'}^{[l']}(c[p])$;
part~($a$) follows. 
If $TR_{i',p}^{[l']}$ finds $r_{i',p}\not=w_{p,i'}$ on
$R_{i'}^{[l',2]}(w_{p,i'})$, then part~($b$) follows by Lemma~\ref{lem5}.
Assume $TR_{i',p}^{[l']}$ finds $r_{i',p}=w_{p,i'}$ on
$R_{i'}^{[l',2]}(w_{p,i'})$.
 From the above two cases, we have $TW_{p}^{[k_{j-1}]}\ra R_{i'}^{[l',1]}(w_{p,i'})$.
Then part~($b$) follows by Lemmas~\ref{lem4} and \ref{lem6}.
$\Box$

\begin{theorem}\label{theo1}
The construction of Figure~\ref{fig1} is a correct implementation of wait-free
bounded concurrent timestamp systems.
\end{theorem}
\noindent{\em Proof:\/}
The wait-freedom property is immediate from the structure of the four
routines in Figure~\ref{fig1}.
The boundedness follows from Corollary~\ref{cor2}.
We now show that the construction satisfies all the four properties P1--P4
described in Section~\ref{sec2}.

\noindent{\em Ordering:} 
Consider two labeling operation executions $L_p^{[k]}$ and $L_{q}^{[k']}$
with labels $l_p^{[k]}$ and $l_{q}^{[k']}$, respectively.
Let $m$ be the least significant index such that
$l_p^{[k]}[m]\not=l_{q}^{[k']}[m]$.
Assume these private values $l_p^{[k]}[m]$ and $l_{q}^{[k']}[m]$ are written by $P_m$ at labeling operation executions
$L_m^{[s_o]}$ and $L_m^{[s_{o'}]}$, respectively.
We define $L_p^{[k]}\Rightarrow L_{q}^{[k']}$ {\em iff\/} $L_m^{[s_o]}\ra L_m^{[s_{o'}]}$.
\begin{itemize}
\item Precedence: Without loss of generality we assume $L_p^{[k]}\ra L_{q}^{[k']}$.
By Lemmas~\ref{lem5} and \ref{lem6}, we have $\pi(TR_{p,m}^{[k]})$ is
$TW_m^{[s_o]}$ and  $\pi(TR_{q,m}^{[k']})$ is $TW_m^{[s_{o'}]}$.
Then, from $TR_{p,m}^{[k]}\ra TR_{q,m}^{[k']}$ and Lemma~\ref{lem9}($b$), we have
$s_{o'}\geq s_o$. As $l_p^{[k]}[m]\not=l_{q}^{[k']}[m]$, we have $s_{o'}\not=s_o$,
and hence, $s_{o'}>s_o$.
That is, $L_m^{[s_o]}\ra L_m^{[s_{o'}]}$.
The precedence property follows.

\item Consistency: For any two labels $l_p^{[k]}$ and $l_{q}^{[k']}$
(returned by a Scan)
such that $m$ is the least significant index for which
$l_p^{[k]}[m]\not=l_{q}^{[k']}[m]$. 
We define $l_p^{[k]}\prec l_{q}^{[k']}$ iff $l_p^{[k]}[m]\prec_m
l_{q}^{[k']}[m]$ iff $L_m^{[s_o]}\ra L_m^{[s_{o'}]}$.
The consistency property follows by Lemma~\ref{lem8} and the definition of 
$\Rightarrow$ given above.
\end{itemize}

\noindent{\em Regularity:} Consider a Scan $S_i^{[j]}$ that returns a label $l_p^{[m_o]}$ that is written by a labeling operation execution $L_p^{[m_o]}$, that
is, $\pi(TR_{i,p}^{[j]})$ is $TW_p^{[m_o]}$.
By Lemmas~\ref{lem5} and \ref{lem6}, we can say $TW_p^{[m_o]}\da
TR_{i,p}^{[j]}$, and hence, $L_p^{[m_o]}\da S_i^{[j]}$.
The second part of the regularity property follows from:
(i)~if $TR_{i,p}^{[j]}$ finds $\rpi\not=\wpi$ on $R_{i}^{[j,2]}(\wpi)$, then,
by Lemma~\ref{lem5},
$\pi(TR_{i,p}^{[j]})$ is $TW_p^{[m_o]}$, where $\pi(R_{i}^{[j]}(c[p]))$ is
$W_p^{[m_o]}(c[p])$, and so,
$TW_p^{[m_{o+1}]}\not\ra TR_{i,p}^{[j]}$, and hence $L_p^{[m_{o+1}]}\not\ra
S_i^{[j]}$;
(ii)~if $TR_{i,p}^{[j]}$ finds $\rpi=\wpi$ on $R_{i}^{[j,2]}(\wpi)$, then,
by Lemma~\ref{lem6},
$\pi(TR_{i,p}^{[j]})$ is $TW_p^{[m_o]}$, where 
$R_{i}^{[j,1]}(\wpi) \LA W_{p}^{[m_o]}(\wpi)\LA R_{i}^{[j,2]}(\wpi)$, and  so,
$TW_p^{[m_{o+1}]}\not\ra TR_{i,p}^{[j]}$, and hence $L_p^{[m_{o+1}]}\not\ra
S_i^{[j]}$.

\noindent{\em Monotonicity:}
Consider two Scans $S_i^{[j]}\ra S_{i'}^{[j']}$. 
Let $S_i^{[j]}$ return label $l_p^{[m_o]}$ from a process $P_p$.
By Lemmas~\ref{lem5} and \ref{lem6}, we have $\pi(TR_{i,p}^{[j]})$ is
$TW_p^{[m_o]}$.
 From $S_i^{[j]}\ra S_{i'}^{[j']}$, we have $TR_{i,p}^{[j]}\ra TR_{i',p}^{[j']}$.
The monotonicity property follows by Lemma~\ref{lem9}.

\noindent{\em Extended regularity:} 
Consider a Scan $S_i^{[j]}$ that returns a label $l_p^{[m_o]}$ that is 
written by a labeling operation execution $L_p^{[m_o]}$, that
is, $\pi(TR_{i,p}^{[j]})$ is $TW_p^{[m_o]}$.
For each labeling operation execution $L_q^{[m']}$, if $S_i^{[j]}\ra
L_q^{[m']}$, then $TR_{i,p}^{[j]}\ra TR_{q,p}^{[m']}$.
Then, by Lemma~\ref{lem9}($a$), we have 
$W_{p}^{[m_o]}(c[p])\LA R_q^{[m']}(c[p])$ and hence, $\pi(TR_{q,p}^{[m']})$ is
$TW_p^{[m_o]}$ or a successor, by  Lemma~\ref{lem9}($b$).
Also by Lemma~\ref{lem5} and \ref{lem6} and the LABELING procedure, we have
$TR_{p,s}^{[m_o]}\ra TW_{p}^{[m_o]}\da TR_{i,p}^{[j]}\ra TR_{q,s}^{[m']}$
for all $s\not=p$,
that is, $TR_{p,s}^{[m_o]}\ra TR_{q,s}^{[m']}$.
Hence, $L_q^{[m']}$ reads more recent (at least equal) private values of all 
processes than $L_p^{[m_o]}$. Also, we have $l_p^{[m_o]}[q]\prec_q l_q^{[m']}[q]$.
Hence $L_p^{[m_o]}\Rightarrow L_q^{[m']}$.
The extended regularity property follows.~$\Box$

\section{Concluding Remarks} \label{sec5}
This paper combines the preliminary \cite{vit86,hald93}.
The former paper is the first to characterize multiwriter shared variables, and
provides a bounded construction of multiwriter multireader multivalued
atomic variable from 1-writer variables. However, it was later found that the 
proposed construction doesn't satisfy 
some properties of atomic shared variables \cite{vit87}.
The technical report \cite{hald93} corrected and extended \cite{vit86} to a construction
of a concurrent timestamp system using an idea from 
\cite{dwork92}. The final result is very close
to the incorrect construction of \cite{vit86}.
It uses $O(n\log n)$ bit size shared variables ($order$ and
$lend$ variables), where $n$ is
the number of processes. Scan and labeling operation executions require $O(n)$
steps.
The construction uses less shared space than that of \cite{dwork92}
at the fundamental level, and is orders of magnitude 
more efficient in terms of scanning bits at the fundamental level.

\subsection{Comparison with Related Work}
In \cite{dwork92}, they have defined three routines, namely,
traceable-read, traceable-write and garbage collection. When the
traceable-read function is executed to read a
label,  the executing process explicitly informs the other processes which of
their private values it is going to use.
The traceable-write procedure is executed to write a new label.
To determine which of its private values are currently in use, a process
executes the garbage collection routine. This routine helps processes to
safely recycle their respective private values.
This is the most intricate routine.

In our construction, we have used a separate implementation technique for
a weaker form of the traceable-read and the traceable-write routines. We do not need a garbage collection
routine.
When a process executes the traceable-read function, it does not explicitly
inform the other processes which of their private values it is going to use.
On the other hand, the executers of the traceable-write procedure correctly finds which
private values of which processes are in use.

Every process needs a separate pool of private values, whose size is
fewer than $2n^2$.
In their construction, the pool size is $22n^2$.
All the ordering shared variables used in our construction are of 1-writer
1-reader regular ones, whereas they are 1-writer $n$-reader atomic ones in their construction.
In our construction, a Scan reads at most $n-1$ 1-writer 1-reader regular ordering shared variables,
whereas in their construction it is $2n-2$ 1-writer $n$-reader atomic ones.
In our construction all but one bit are nonatomic 1-writer 1-reader variables.
Table~\ref{tab1} presents some comparison results briefly.
\begin{table}[h]
\caption{Comparison Results.}\label{tab1}
\vspace{2mm}
\center{
\begin{tabular}{|c|c|c|c|c|c|} \hline
Construction&Shared variable size & Shared space(bits) & Labeling & Scan \\ \hline
\cite{dole89} & $O(n)$ & $O(n^3)$ & $O(n)$  & $O(n^2\log n)$  \\ \hline
\cite{gawl92} & $O(n^2)$ & $O(n^4)$ & $O(n\log n)$  & $O(n\log n)$  \\ \hline
\cite{isra92} & $O(n^2)$ & $O(n^4)$ & $O(n)$  & $O(n)$  \\ \hline
\cite{dwork92} & $O(n\log n)$  & $O(n^5\log n)$ & $O(n)$  & $O(n)$  \\ \hline
\cite{dwork92a} & $O(n)$  & $O(n^3)$ & $O(n)$  & $O(n)$  \\ \hline
This paper & $O(n\log n)$ & $O(n^3\log n)$ & $O(n)$  & $O(n)$ \\ \hline
\end{tabular}
}
\end{table}

Of all proposed constructions of bounded concurrent timestamp systems
we are aware of, the construction in this paper is the `simplest'.
The correctness proof, though involved, is easier to follow.
It is used as a basis in the reference text \cite{atti99} to describe 
bounded concurrent timestamp system.

Although we have used a notion of vector clocks for our construction, 
as in \cite{vit86}, we may not
really need the full power of vector clock concept developed 
later by Mattern \cite{matt89}.
In CTSs, we are not interested in determining causal `independence'
of various labeling operation executions.
The ordering property of CTSs infers that the causal orders among
labeling operation executions matter most.
We need to have a total order on all labeling operation executions,
and the total order must extend their original causal relation.
This is akin to the logical time of Lamport \cite{lamp78}.
We suspect that there might be a way to eliminate the 
vector clock altogether, by an efficient way of recycling of 
global values, instead of using $n$ sets of private values.

The construction presented here should not be considered as an alternative implementation of
the traceable use abstraction, for it restricts the value propagation at indirection level one.
It is not clear to the authors how this strategy could be extended for a general 
implementation of the abstraction.
%Also, it will be interesting to see some strategy to recycle global
%values rather than private ones.

\subsection{A Brief Early History}
The development of bounded wait-free shared variables and timestamp
systems has been quite problematic and error-prone.
It may be useful at this point to present a brief early history of the area:
who did what, when, and where, and which solutions are known to be incorrect.
In a series of papers \cite{lamp74,lamp77,lamp78,lamp86} starting
in 1974, Lamport
explored various notions of concurrent reading and writing of shared
variables culminating in the seminal 1986 paper \cite{lamp86}. It formulates
the notion of wait-free implementation of an atomic shared variable---written
by a single writer
and read by (another) single reader---from safe 1-writer 1-reader
2-valued shared variables, being mathematical versions of physical 
{\em flip-flops}. Predating the latter
paper, in 1983 Peterson
\cite{pete83} published an ingenious wait-free construction of an atomic
1-writer, $n$-reader $m$-valued atomic shared variable from
$n+2$ safe 1-writer $n$-reader $m$-valued registers, $2n$ 
1-writer 1-reader 2-valued
atomic shared variables, and 2 1-writer $n$-reader 
2-valued atomic shared
variables. He presented also a proper notion of wait-freedom property.
Lamport \cite{lamp84} gave an example that appeared to contradict
a possible interpretation of the informal statement of
a theorem in \cite{pete83},
which, as Peterson apparently retorted to Lamport, was not intended. 
In his paper, Peterson didn't tell how to construct the $n$-reader
boolean atomic variables from flip-flops, while Lamport mentioned the open
problem of doing so, and, incidentally, uses a version
of Peterson's construction to bridge the algorithmically demanding
step from atomic shared bits to atomic shared multivalues.
Based on this work, N. Lynch, motivated by concurrency control of
multi-user data-bases, posed around 1985 the question of how to
construct wait-free multiwriter atomic variables from
1-writer multireader atomic variables. Her student Bloom
\cite{bloom87} found in 1985 an elegant 2-writer construction, which,
however, has resisted generalizations to multiwriter. Vit\'anyi and Awerbuch
\cite{vit86} were the first to define and explore the complicated notion of 
wait-free constructions of general
multiwriter atomic variables. They presented
a proof method, an unbounded solution 
from 1-writer 1-reader atomic variables, and a bounded solution
from 1-writer $n$-reader atomic variables. The unbounded solution
was made bounded in \cite{li89}. It is optimal for the implementation
of $n$-writer $n$-reader atomic variables from 1-writer 1-reader ones.
``Projections'' of the construction also give specialized constructions
for  the implementation of $1$-writer $n$-reader atomic 
variables from $1$-writer $1$-reader ones, and for the implementation
of $n$-writer $n$-reader atomic variables from 1-writer $n$-reader ones. 
As noted in \cite{LV92},
the first ``projection'' is optimal, while the
last ``projection'' may not be optimal since it uses $O(n)$
control bits per writer while only a lower bound of $\Omega (\log n)$
was established. Taking up this
challenge, the construction in \cite{isra92a} apparently achieves this
lower bound. The earlier bounded solution in \cite{vit86} (corresponding
in fact to the problem correctly solved by the last ``projection'' above) 
turned out
not to be atomic, but only achieved regularity \cite{vit87}.
Nonetheless, \cite{vit86} introduced important notions and technique in the
area, like (bounded) vector clocks. These were inspired by the
celebrated ``Bakery'' algorithm of Lamport \cite{lamp74}, which can
be viewed as a
global bounded ``clock'' determining the order among queued processes
much like the ticket dispenser in a bakery serves to determine
the order of servicing waiting customers. The multiwriter situation
has stronger requirements than apparently
can be satisfied by a global ticket dispenser.
The solution in \cite{vit86} was the construction 
of a bounded ``vector clock'':
a private ticket dispenser
for each process, the storing and updating 
of a vector of latest tickets
held by all processes, together with a semantics to determine the order
between vectors. Moreover, a complex mechanism---primitive traceable
read/write---is presented to keep track
of which tickets of what processes could still be present in the system,
with the objective of bounding the private ticket pool of
each process by recycling obsolete tickets. Following
the appearance of \cite{vit86}, 
Peterson who had been working on the multiwriter problem
for a decade,
% (letter Oct. 86 to the second author), 
together with Burns, revamped the construction
retaining the vector clocks, but replaced the primitive traceable
read/write elements by repeated scanning as in \cite{pete83}. The result
\cite{pete87} was found to be nonetheless erroneous, in the technical
report \cite{scha88}. This makes the multiwriter problem
perhaps the only one for which two consecutive wrong solutions were
published in the highly selective FOCS conferences. Neither the 
re-correction in \cite{scha88}, nor the claimed re-correction by
the authors of \cite{pete87} has appeared in print.
The present paper constitutes a correction of the original \cite{vit86}
by the extension of \cite{hald93}: by implementing the stronger
concurrent timestamp system it also solves the atomic multiwriter problem.
Apart from the already mentioned \cite{li89},
the only other multiwriter multireader atomic shared variable construction
that appeared in journal version seems to be of Abraham \cite{abra95}.
Also in 1987 there appeared at least five purported solutions for the
implementation of 1-writer $n$-reader atomic shared variable from
1-writer 1-reader ones: \cite{kiro87,wolf87,burn87,sing87} and
the conference version of \cite{isra93},
of which \cite{burn87} was shown to be incorrect in \cite{hald92}
and only \cite{sing87} appeared in journal version.
The only other 1-writer $n$-reader atomic shared variable construction
appeared in journal version is of Haldar and Vidyasankar \cite{hald91}.
A. Israeli and M. Li were attracted to the area by the work in \cite{vit86},
and, in an important paper \cite{isra93}, they raised and solved the question
of the more general and universally useful
notion of bounded timestamp system to track the order
of events in a concurrent system. Their sequential timestamp
system was published in journal version, but the preliminary concurrent
timestamp system in the conference proceedings, of which a more
detailed version has been circulated in manuscript form,
has not been published in final form. 

The difficulty of wait-free atomic multireader-, multiwriter-,
and timestamp system constructions, 
and the many errors in purported 
and published solutions, have made it hard to publish results in print.
Of the major pioneering papers, the first correct multiwriter
construction of 1987 \cite{li89} was rejected at five
consecutive conferences until it was published in ICALP, 1989. The final journal
version was handled by three consecutive editors, scrutinized by three
consecutive sets of referees, and lasted from 1989 until publication in 1996.
The pioneering timestamp paper, \cite{isra93}, was submitted in 1987/88 to
this journal, after a couple of years rejected since a stronger result
% Haldar: how do we know about this rejection????
\cite{dole89} had appeared in conference version, submitted to another
journal and finally appeared in 1993, but only the part containing
the simpler sequential timestamp construction. The first
generally accepted concurrent timestamp construction \cite{dole89}
appeared in conference version in 1989, but its journal version appeared
only in 1997. As stated before, the concurrent timestamp construction in the
present paper is based on the 1986 paper \cite{vit86} 
supplemented by the 1993 technical report \cite{hald93}.
For further remarks see \cite{li89}
in this journal and the Introduction to present paper.

\section*{Acknowledgment}
Hagit Attiya and Orli Waarts gave valuable suggestions
for an early version of \cite{hald93} and Baruch Awerbuch
co-authored the old preliminary paper \cite{vit86} on which this 
paper is based.

\newpage 
\setlength{\textheight}{9.00in}
\setlength{\textwidth}{6.50in}
\setlength{\footheight}{0.0in}
\setlength{\topmargin}{-1cm}
\setlength{\headheight}{0.0in}
\setlength{\headsep}{0.0in}
\setlength{\oddsidemargin}{-10mm}
\setlength{\parskip}{1ex plus .2ex minus .2ex}
\newcommand{\cl}{cl}

\begin{figure}%fig1
\begin{tabbing}
De\=clarations \kill 
{\bf Declarations}\\
\\
\>Co\=nstants: \\[2mm]
\>\>$n$ = number of processes; \\
\\
\>Type:\\[2mm]
\>\>label-type: array  [1..$n$] of natural number; \= \{represents vector clock\}\\
\>\>boolean: 0..1;\\
\\
\>Sh\=ared variables and their initial values: \\[2mm]
\>\>\WC\ : array [1..$n$, 1..$n$] of boolean {\em atomic\/}; \>\{{\bf all initially $0$}\} \\
\>\>\>\{$P_p$ writes $\WC[p,1..n]$ and ~$P_i$ reads $\WC[1..n,i]$\} \\
\mbox{} \\
\>\>\RC\ : array [1..$n$, 1..$n$] of boolean {\em atomic\/}; \>\{{\bf all initially $0$}\} \\
\>\>\>\{$P_p$ writes $\RC[p,1..n]$ and ~$P_i$ reads $\RC[1..n,i]$\}\\
\mbox{} \\
\>\> $c$ :  array [1..$n$] of boolean {\em atomic\/};  \>\{{\bf initially} 0\} \\
\>\>\>\{$P_p$ writes $c[p]$, and the others read\} \\
\mbox{} \\
\>\> $label$ : array [1..$n$, 0..1] of label-type {\em safe\/}; \>\{{\bf all
initially $0$, except $label[p,0][p]=1$ for all $p$}\}\\
\>\>\>\{$P_p$ writes $label[p,0..1]$ and the others read\} \\
\mbox{} \\
\>\>$copylabel$ : array [1..$n$, 1..$n$] of label-type {\em safe\/}; \\
\>\>\>\{$P_p$ writes $copylabel[p,1..n]$ and $P_i$ reads $copylabel[1..n,i]$\} \\
\mbox{} \\
\>\>$lend$: array [1..$n$, 1..$n$] of {\em regular\/} array [0..1] of label-type; \{{\bf all initially} 0\} \\
\>\>\>\{$P_p$ writes $lend[p,1..n]$ and $P_i$ reads $lend[1..n,i]$\} \\
\mbox{} \\
\>\>$order$: array [1..$n$, 1..$n$] of {\em regular\/} array [1..$5n$] of natural number; \\
\>\>\>\{{\bf initially} $order[1..n,1..n][1]=0$ and $order[1..n,1..n][2]=1$\}\\
\>\>\>\{$P_p$ writes $order[p,1..n]$ and $P_i$ reads $order[1..n,i]$\} \\
\mbox{} \\
\mbox{} \\
\>Private variables for process $P_p$, $p=1,2,\ldots,n$:\\[2mm]
\>\>$\cl_p$: boolean; \>\{{\bf initially} 0\} \\
\>\>$myLend_p$: array $[1..n$] of array [0..1] of label-type; \{{\bf all initially 0}\}\\ 
\>\>$old$-$label_p$: label-type;  \>\{{\bf all initially} 0, except
$old$-$label_p[p]=1$\} \\
\>\>$\prec_p$: total order relation; \>\{{\bf initially} \{$\langle 0,1\rangle\}$\} \\
\end{tabbing}
\caption{Shared variables.} 
\end{figure}

\addtocounter{figure}{-1}

\begin{figure}
\begin{tabbing}
{\bf Procedure} TRACEABLE-WRITE($p$: 1..$n$; $new$-$label$: label-type); \{$P_p$ writes $new$-$label$ in ${\cal O}[p]$\}\\
var \= \\
\>$i,j$: 1..$n$; \{loop index\}\\
\>$lr$: boolean; \\
begin \\
1.\>$\cl_p := \neg \cl_p$; \\ 
2.\>write $new$-$label$ in $label[p, \cl_p]$; \\
3.\>write $\cl_p$ in $c[p]$; \\
4.\>for \= $i := 1$ to $n$ do\=\= \\
\>\>be\=gin  \>\hspace{4cm}\=\{{\em could be done in parallel\/}\} \\
4.1\>\>\>read $lr$ from $\RC[i,p]$; \\
4.2\>\>\>if \= $lr \not= \WC[p,i]$ then \\
4.2.1\>\>\>\> write $new$-$label$ in $copylabel[p,i]$; \\
4.2.2\>\>\>\> for $j := 1$ to $n$ do $myLend_p[j][0..1][i] := \langle old$-$label_p[j]$, $new$-$label[j]\rangle$; \\
4.2.3\>\>\>\>write $lr$ in $\WC[p,i]$; \> \{$\WC[p,i]=\RC[i,p]$\} \\
\>\>\>endif; \\
\>\>endfor; \\
5.\>for $j := 1$ to $n$ do $myLend_p[j][1][p]:= new$-$label[j]$; \\
6.\>for $j := 1$ to $n$ do write $myLend_p[j]$ in $lend[p,j]$; \{{\em could be done in parallel\/}\}\\
7.\> $old$-$label_p :=new$-$label$;\\
end; \{of procedure\}\\
\mbox{} \\
{\bf Function} TRACEABLE-READ($p$: 1..$n$, $i$: 1..$n$): label-type;  \{$P_p$ reads a label from $P_i$\}\\
var \= \\
\>$lw$: boolean; \\
\>$lc$: boolean; \\
\> $savelabel$: label-type; \\
begin \\
1.\>read $lw$ from $\WC[i,p]$;  \\
2.\>write $\neg lw$ in $\RC[p,i]$; \>\>\>\>\{$\RC[p,i]\not=\WC[i,p]$\} \\
3.\>read $lc$ from $c[i]$; \\
4.\>read $savelabel$ from $label[i, lc]$; \\
5.\>read $lw$ from $\WC[i, p]$; \\
6.\>if \=$(\RC[p,i] \not= lw)$ then return($savelabel$) \\
7.\>else \>\>\>\>\{$\RC[p,i] = \WC[i,p]$\} \\
\>\>read and return($copylabel[i,p]$) \\
\>endif;\\
end; \{of function\}
\end{tabbing}
\caption{Construction for process $P_p$. (Cont'd.)} 
\end{figure}

\addtocounter{figure}{-1}

\begin{figure}
\begin{tabbing}
{\bf Procedure} LABELING($p$: 1..$n$); \\
var \= \\
\>$j,k$: $1..n$; \\
\>$temp$: array $[1..n]$ of array [0..1] of label-type; \\
\>$lab$: array $[1..n]$ of label-type; \\
\>$new$-$label$: label-type; \\
\>$private$-$value$: natural number; \\
begin \\
1.\> for \=$j := 1$ to $n$ do \{{\em could be done in parallel\/}\} \\
\>\> read $temp[j]$ from $lend[j,p]$; \{{\bf we do not need
$temp[j][0][j]$}\} \\
2.\> select a new $private$-$value$ not in $temp[1..n]$ and the current private value; \{use the axiom of choice here\}\\
3.\> put the new $private$-$value$ in $\prec_p$ as the largest element; \\
4.\> for \=$j := 1$ to $n$ do \{{\em could be done in parallel\/}\} \\
4.1\>\> order the elements of (\=$temp[1..n][0..1][j]$, \\
\>\>\>                       $temp[k][1][k]$ and \\
\>\>\>                       $temp[p][0..1][k]$ for all $k$, \\
\>\>\> and the new $private$-$value$) consistent with $\prec_p$ \\
4.2\>\> and write them in $order[p,j]$; \\
5.\> for $j := 1$ to $n$, $j\not=p$, do $lab[j]:=$TRACEABLE-READ($p,j$); \{{\em could be done in parallel\/}\}\\
6. \> $new$-$label$ := $\langle lab[1][1], lab[2][2], \dots, lab[p][p]:=private$-$value, \dots, lab[n][n]\rangle$; \\
7.\>TRACEABLE-WRITE($p$, $new$-$label$);\\
end;\\
\mbox{}\\
{\bf Function} SCAN($p$: 1..$n$):$(\overline{l}, \prec)$; \\
var \= \\
\>$i,j,k$: $1..n$; \\
\>$lab$: array $[1..n$] of label-type;\\
begin \\
1.\> for $j := 1$ to $n$ do $lab[j]:=$TRACEABLE-READ($p,j$); \{{\em could be done in parallel\/}\}\\
2.\> for \=$i:= 1$ to $n$ do \\
2.1\>\> for \=$j := 1$ to $n$ do \\
2.1.1\>\>\> let $k$ be the least significant index in which $lab[i]$ differs from
$lab[j]$; \\
2.1.2\>\>\> if $order[k,p]$ (which is a subset of $\prec_k$) is not read yet, then read it; \\
2.1.3\>\>\> determine the order between $lab[i]$ and $lab[j]$ using  $\prec_k$; \\
end;
\end{tabbing}
\caption{Construction for process $P_p$. (Cont'd.)} \label{fig1}
\end{figure}

\end{document}